\documentclass[fleqn,12pt,twoside]{article}
\usepackage[headings]{espcrc1}

\readRCS
$Id: espcrc1.tex,v 1.2 2004/02/24 11:22:11 spepping Exp $
\usepackage{graphicx}
\usepackage[figuresright]{rotating}

\newcommand{\AmS}{{\protect\the\textfont2
  A\kern-.1667em\lower.5ex\hbox{M}\kern-.125emS}}

\title{Color superconductivity in a strong external magnetic field}

\author{Cristina Manuel\\
Instituto de Ciencias del Espacio (IEEC/CSIC), Campus UAB, Fac. de Ci\`encies,\\
Torre C5-Parell 2a planta, E-08193, Bellaterra (Barcelona) Spain}%
\runtitle{Color superconductivity in a strong B field}
\runauthor{C. Manuel}

\begin{document}

% typeset front matter
\maketitle

\begin{abstract}
We explore
the effects of an applied strong external magnetic field in a three flavor
massless color superconductor. The long-range component of the B field
that penetrates the superconductor enhances some quark condensates,
leading to a different condensation pattern. The external field also
reduces the flavor symmetries in the system, and thus it changes drastically
the  corresponding low energy physics. Our considerations are
relevant for the study of highly magnetized compact stars.

\end{abstract}

\section{Introduction}

In the interior of some compact stars the density is so high that the hadrons melt into
their fundamental constituents,
giving rise to quark matter. It has been known for long time now that cold dense quark matter should exhibit the phenomenon of
color superconductivity \cite{reviews}. It is our aim here to explain how a strong magnetic
field affects this phenomenon. This is not
a simple academic question.
The real fact is that almost all compact stars sustain a strong magnetic field, of the order of $B \sim 10^{12} - 10^{14}$ G for pulsars,
and of $B \sim 10^{14}-10^{15}$ G for magnetars. A comparison of the gravitational and magnetic energies of a compact star tells us that the
maximum fields may be as high as $B \sim 10^{18}-10^{19}$ G.
The common believe is that all the above mentioned compacts objects are neutron
stars, where neutrons are in a superfluid phase, while the protons are in a superconducting one, probably
of type I.
An external magnetic field influences both quantitatively
and qualitatively the color superconducting state \cite{MCFL}. We will
 explain here why this is so. This work has been done in collaboration with Efrain J. Ferrer and
Vivian de la Incera.

\section{Color-flavor locking phase}

The ground state of QCD at high baryonic density with three
light quark flavors is described by the (spin zero) condensates \cite{alf-raj-wil-99/537}
\begin{equation}
  \langle q^{ai}_{L } q^{bj}_{L }\, \rangle
=-\langle q^{ai}_{R } q^{bj}_{R }\, \rangle
=   \Delta_A \, \epsilon^{abc} \epsilon_{ijc}
\ ,
\end{equation}
where $q_{L/R}$ are Weyl spinors (a sum over spinor indices is
understood), and $a,b$ and $i,j$ denote flavor and color indices,
respectively. For simplicity we have neglected a color sextet component of the condensate,
as it is a subleading effect.

The diquark condensates lock the color and flavor transformations
breaking both- thus, the name color-flavor locked (CFL) phase.
 The symmetry breaking pattern  in the CFL phase is
$
 SU(3)_C \times SU(3)_L \times SU(3)_R \times U(1)_B
\rightarrow SU(3)_{C+L+R}$.
There are only nine Goldstone bosons that survive to the
Anderson-Higgs mechanism. One is a singlet, scalar mode,
associated to the breaking of the baryonic symmetry, and the
remaining octet is associated to the breaking of the axial $SU(3)_A$ group, just
like the octet of mesons in vacuum.

An important feature of spin-zero color superconductivity is that
although the color condensate has non-zero electric charge, there is
a linear combination of the photon $A_{\mu}$ and a gluon
$G^{8}_{\mu}$ that remains massless \cite{alf-raj-wil-99/537},
$
\widetilde{A}_{\mu}=\cos{\theta}\,A_{\mu}-\sin{\theta}\,G^{8}_{\mu} \ ,
$
while the orthogonal combination
$\widetilde{G}_{\mu}^8=\sin{\theta}\, A_{\mu}+\cos{\theta}\,G^{8}_{\mu}$
is massive. In the CFL phase the mixing angle $\theta$ is
sufficiently small ($\sin{\theta}\sim e/g\sim1/40$). Thus, the
penetrating field in the color superconductor is mostly formed by
the photon with only a small gluon admixture.

The unbroken $U(1)$ group corresponding to the long-range rotated
photon (i.e. the $\widetilde {U}(1)_{\rm e.m.}$) is generated, in
flavor-color space, by $\widetilde {Q} = Q \times 1 - 1 \times Q$,
where $Q$ is the electromagnetic charge generator. We use the
conventions $Q = -\lambda_8/\sqrt{3}$, where $\lambda_8$ is the 8th
Gell-Mann matrix. Thus our flavor-space ordering is $(s,d,u)$. In
the 9-dimensional flavor-color representation that we will use here
(the color indexes we are using are (1,2,3)=(b,g,r)), the
$\widetilde{Q}$ charges of the different quarks, in units of
$\widetilde{e} = e \cos{\theta}$, are
\begin{equation}
\label{q-charges}
\begin{tabular}{|c|c|c|c|c|c|c|c|c|}
  \hline
  % after \\: \hline or \cline{col1-col2} \cline{col3-col4} ...
  $s_{1}$ & $s_{2}$ & $s_{3}$ & $d_{1}$ & $d_{2}$ & $d_{3}$ & $u_{1}$ & $u_{2}$ & $u_{3}$ \\
  \hline
  0 & 0 & - & 0 & 0 & - & + & + & 0 \\
  \hline
\end{tabular}
\end{equation}

While a weak magnetic field only changes slightly the properties of the CFL superconductor,
in the presence of a strong magnetic field the condensation pattern is changed,
giving rise to a new phase, the magnetic color-flavor locked (MCFL) phase.

\section{Magnetic color-flavor locking phase}

An external magnetic field to the color superconductor will be able to penetrate it in
the form of a ``rotated" magnetic field $\widetilde{B}$. With respect to this long-ranged field,
although all the superconducting pairs are neutral, a subset of them are formed by quarks with
opposite rotated $\widetilde{Q}$ charges.  Hence, it
is natural to expect that this kind of condensates will be
affected by the penetrating field, as the quarks couple minimally to the rotated
gauge field. Furthermore, one may expect that these condensates are
strengthened by the penetrating field, since their paired quarks,
having opposite $\widetilde{Q}$-charges and opposite spins, have
parallel (instead of antiparallel) magnetic moments.
 The situation
here has some resemblance to the magnetic catalysis of chiral
symmetry breaking \cite{MC}, in the sense that the magnetic field
strengthens the pair formation. Despite this similarity, the way the
field influences the pairing mechanism in the two cases is quite
different as we will discuss later on.

A strong magnetic field affects the flavor symmetries of QCD, as
different quark flavors have different electromagnetic charges.
For three light quark flavors, only the subgroup of $SU(3)_L
\times SU(3)_R$ that commutes with $Q$, the electromagnetic
generator, is a symmetry of the theory. Equally, in the CFL phase
a strong $\widetilde{B}$ field should affect the symmetries in the
theory, as $\widetilde{Q}$ does not commute with the whole locked
$SU(3)$ group. Based on this considerations, we proposed the
following diquark (spin zero) condensate \cite{MCFL}
\begin{equation}
  \langle q^{ai}_{L } q^{bj}_{L }\, \rangle
=-\langle q^{ai}_{R } q^{bj}_{R }\, \rangle
=   \Delta_A \, \epsilon^{ab3} \epsilon_{ij3} + \Delta_A^B \left( \epsilon^{ab2} \epsilon_{ij2} +
\epsilon^{ab1} \epsilon_{ij1} \right)
\ ,
\end{equation}
and as for the CFL case, we have only considered the leading antritiplet color channel.
For a discussion of the remaining allowed structures in the subleading sextet channel see \cite{MCFL}.
Here we have been guided by the principle of highest symmetry, that is, the pair
condensation should retain the highest permitted degree of symmetry,
as then quarks of different colors and flavors will participate in
the condensation process to guarantee a maximal attractive channel
at the Fermi surface \cite{alf-raj-wil-99/537}.

In the MCFL phase
the symmetry breaking pattern is
$
SU(3)_C \times SU(2)_L \times SU(2)_R \times U(1)^{(1)}_A\times U(1)_B \times U(1)_{\rm e.m.}
\rightarrow SU(2)_{C+L+R} \times {\widetilde U(1)}_{\rm e.m.}
$.
Here the symmetry group $U(1)^{(1)}_A$ is related to a current
which is an anomaly free linear combination of $u,d$ and $s$ axial
currents, and such that  $U(1)^{(1)}_A \subset SU(3)_A$. The locked $SU(2)$ group
corresponds to the maximal unbroken symmetry, such that it maximizes the condensation
energy. The counting of broken generators, after taking into account the
Anderson-Higgs mechanism, tells us that there are only five Goldstone
bosons. As in the CFL case, one is associated to the breaking of
the baryon symmetry; three Goldstone bosons are associated to the
breaking of $SU(2)_A$, and another one associated to the breaking
of  $U(1)^{(1)}_A$.

To study the MCFL phase we used a Nambu-Jona-Lasinio (NJL)
four-fermion interaction abstracted from one-gluon exchange
\cite{alf-raj-wil-99/537}. This simplified treatment, although
disregards the effect of the $\widetilde {B}$-field on the gluon
dynamics and assumes the same NJL couplings for the system with
and without magnetic field, keeps the main attributes of the
theory, providing the correct qualitative physics.  The NJL
model is treated as the proper effective field theory to study color
superconductivity in the regime of moderate densities.
The model is defined by two parameters, a coupling constant
$g$ and an ultraviolet cutoff $\Lambda$. The cutoff should be much
higher than the typical energy scales in the system, that is, the
chemical potential $\mu$ and the magnetic energy
$\sqrt{\widetilde{e}\widetilde{B}}$.

The  MCFL gap equations for arbitrarily value of the magnetic field
are extremely difficult to solve, and they require a numerical treatment.
However, we have found a situation where an analytical solution can
be found. This corresponds to the case
 $\widetilde{e}\widetilde{B} >\mu^2/2$, where $\mu$ is the chemical potential. In this case, only
charged quarks in the lowest Landau level contribute to the gap equation,
a situation that drastically simplifies the analysis.

In BCS theory, and in the presence of contact interactions, the fermionic gap has an
exponential dependence on the inverse of the density of states close to the
Fermi surface, which is proportional to $\mu^2$.
Effectively, one can find that within the NJL model the CFL gap reads

\begin{equation}
\label{gapCFL}
\Delta^{\rm CFL}_A \sim 2 \sqrt{\delta \mu} \, \exp{\Big( -\frac{3
\Lambda^2 \pi^2} {2 g^2 \mu^2} \Big) } \ .
\end{equation}
with $\delta \equiv \Lambda - \mu$.
In the MCFL phase, when $\widetilde{e}\widetilde{B} >\mu^2/2$, we find instead
\begin{equation}
\label{gapBA}
\Delta^B_A \sim 2 \sqrt{\delta \mu} \, \exp{\Big( - \frac{3 \Lambda^2
\pi^2} {g^2 \left(\mu^2 + \widetilde{e} \widetilde{B} \right)}
\Big) } \ .
\end{equation}
For the value of the remaining gaps of the MCFL phase, see
\cite{MCFL}. All the gaps feel the presence of the external magnetic
field. As expected, the effect of the magnetic field in
$\Delta^{B}_{A}$ is to increase the density of states, which enters
in the argument of the exponential as typical of a BCS solution. The
density of states appearing in (\ref{gapBA}) is just the sum of
those of neutral and charged particles participating in the given
gap equation (for each Landau level, the density of states around
the Fermi surface for a charged quark is
$\widetilde{e}\widetilde{B}/2 \pi^2$). The gap formed by
$\widetilde{Q}$-neutral particles, although  modified by the
$\widetilde{B}$ field \cite{MCFL}, has a subleading effect in the
MCFL phase.

As mentioned at the beginning of this Section, the situation here
shares some similarities with the magnetic catalysis of chiral
symmetry breaking \cite{MC}; however, the way the field influences
the pairing mechanism in the two cases is quite different. The
particles participating in the chiral condensate are near the
surface of the Dirac sea. The effect of a magnetic field there is
to effectively reduce the dimension of the particles at the lowest
Landau level, which in turn strengthens their effective coupling,
catalyzing the chiral condensate. Color superconductivity, on the
other hand, involves quarks near the Fermi surface, with a pairing
dynamics that is already $(1+1)$-dimensional. Therefore, the
${\widetilde B}$ field does not yield further dimensional
reduction of the pairing dynamics near the Fermi surface and hence
the lowest Landau level does not have a special significance here.
Nevertheless, the field increases the density of states of the
${\widetilde Q}$-charged quarks, and it is through this effect, as
shown in  (\ref{gapBA}), that the pairing of the charged particles
is reinforced by the penetrating magnetic field.

\section{Conclusions}

We have presented the arguments to explain why three light flavor color superconductivity
is made stronger, not weaker, by the presence of magnetism. These arguments have been
corroborated by an explicit computation of the quark gaps within a NJL model, in the regime
of strong magnetic fields. To better understand the
relevance of this new phase in astrophysics we need to explore the
region of moderately strong magnetic fields
$\widetilde{e}\widetilde{B}< \mu^2/2$, which requires to carry out
a numerical study of the gap equations including the effect of
higher Landau levels.

The presence of a strong magnetic field affects the values of the quark
gaps, and thus, it will modify the equation of state of the color superconductor,
although we do not expect this to be a very pronounced effect. More drastically,
the low energy physics of the MCFL phase would differ from that of the CFL phase,
through the disappearance of light degrees of freedom. This fact will have consequences
on several macroscopic properties of the superconductor, that we hope to explore
soon.

\end{document}